# Preprint Imagining In-Air Interaction for Hemiplegia Sufferer

Zhihan Lv
FIVAN, Valencia, Spain
Email: lvzhihan@gmail.com

Haibo Li
Royal Institute of Technology (KTH)
Stockholm, Sweden
Email: haiboli@kth.se

*Abstract*—In this paper, we described the imagination scenarios of a touch-less interaction technology for hemiplegia, which can support either hand or foot interaction with the smartphone or head mounted device (HMD). The computer vision interaction technology is implemented in our previous work, which provides a core support for gesture interaction by accurately detecting and tracking the hand or foot gesture. The patients interact with the application using hand/foot gesture motion in the camera view.

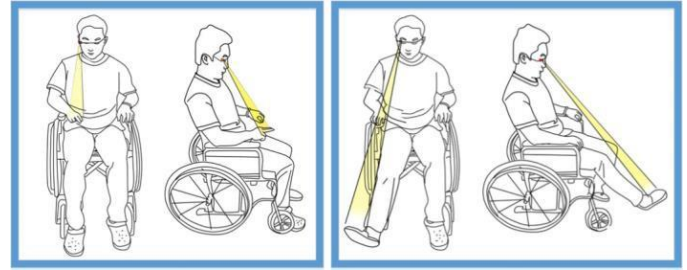

Fig. 1. Imagined scenarios. Left: hand touch-less interaction; Right: foot touch-less interaction.

## I. Introduction

Nowadays, the society appeals all the people to devote sympathy and concern to paralysis sufferers, in which hemiplegia sufferers have the most amount and suffer paralysis of one side of the body. The research about experiencing of being a hemiplegia sufferers is this kind of social contribution [21]. Besides, there have been quite a lot of software and hardware research about virtual rehabilitation tool design and development for hemiplegia currently [4] [2] [8] [5], which are also this kind of contributions for hemiplegia. The purpose of our research is not a rehabilitation assistive tool, however, provides a new low-cost hand and foot interaction approach for hemiplegia. Currently, the related research have been conducted [10] [17] [3] mostly on the specific expensive devices, which cannot be consumed by hemiplegia patients in poor economic condition. The specific game designed for cerebral palsy sufferer are already proved [6] [7], the input device is however designed for common player. A leapmotion based interactive game is designed for stroke sufferer [9].

Recent research in mobile phone computing has led to a number of input techniques that allow for spatial input by hands beyond a touchscreen, all provide a level of spatial input in a mobile setting. Some hemiplegia patients prefer feet since their hands have lost touch of sense. Considering handheld computing various foot gesture detection and tracking system have been proposed such as multitoe, foot tapping. We designed hand and foot gesture interaction on handheld devices or vision based wearable device such as google glasses. The 'anywhere' [18] and 'connectivity' [1] are two marked features of current mobile device which met our practical demand, beside of low-cost.

In this paper, we imagine to utilize our touch-less interaction approach for hemiplegia, where patient wearable vision based head-mounted-device (HMD) and performs 'in air' interaction gestures using fingers or feet. The amount of hemiplegia sufferers is much high. In our clinical sample database, there are sixty hemiplegia sufferers, but no any monoplegia sufferers.

## II. Touch-less Interaction

Four wearable applications to illustrate and explore different use cases (as shown in Figure 2), have presented [11]. The applications are respectively 'Social media site launch' and 'Bouncing ball game' for hand (finger) gesture, as well as 'Football game' and 'Foot-Play Piano' for foot gesture. Accordingly, we report on a user study comparing usability and emotions at VR2015 [13]. We expect the patients to perceive the touch-less approach as being stretched from his/her hand/foot for manipulating content on the virtual scene of the applications and augmented graphic feedback is brought to the vision. In order to complete this interaction loop, our system maps and updates the physical hand/foot onto the glasses projector. The system has been demonstrated [14].

## III. Imagination of Scenarios

The practical scenarios have been presented before. In this paper we imagine that the player is a hemiplegia sufferer who is using the applications, and we will see to what extent the touch-less interaction technology could be used by hemiplegia sufferers.

In the Wearable Circle Menu app, a round-shape menu can lunch several popular social media sites. Patients can rotate the suspended circle menu and select the icon using finger gesture in air. To access the required icon, one can make the circle menu rotating and select any icon. The expected icon can be selected and unselected by the same gesture. The Bouncing ball game is a handball court with a camera view overlay. An augmented reality glove image is rendered following the finger gesture. The handball always bounces back and forth on the court area until it's into the goal or the sufferer intercept and hold it. When the sufferer holds the ball, he can dribble the ball to the desired position by slowly swing finger or throw the ball by swing finger fast. 'Football game' and 'Bouncing ball game' have the nearly same game rulers beside active controlling body are respectively foot and hand. In the Football game, the sufferers use real foot to kick the ball. In the Foot-Play Piano application scenario, a circle progress plate is on the right-top corner of the screen, which is used to show how long the sufferer press the piano key, eg. semi-circle is a half mora. When the volume up button is pressed , the application starts to look for the movement of the foot through camera. When a collision is detected with any of the augmented piano-key, the code related to that key is generated by the program, and the circle progress plate whirls until the sufferer release the piano key.

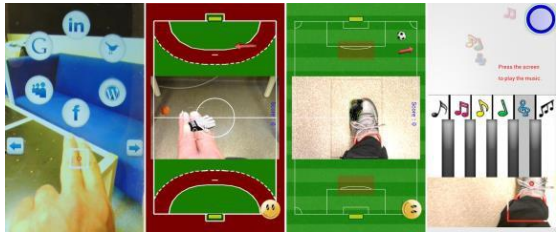

Fig. 2. The previously developed four games based on the interaction approaches. From left-to-right: Wearable circle meanu; Bouncing ball game; Football game; Foot Piano. [11]

## IV. Conclusions

The touch-less interaction solution provides a low-cost solution which is also potentially applicable to wear on other body parts for example tongue. The therapist suggests that the full hand motion is easier for hemiplegia sufferers than finger motion. We believe that it's entirely possible to be equipped on hemiplegia sufferers for their daily interaction. The touch-less interaction technology is also suitable for future rehabilitation application. Touch-less interaction technology has been planed to widely apply in many fields range from everyday entertainment [12] [22] to traditional research fields such as geography [19] [15] [16] and biology [20].

## Acknowledgment

The authors would like to thank Pablo Gagliardo and Sonia Blasco for fruitful discussion and clinical suggestions, and thank Dr. Alaa Halawani and Dr. Shafiq Ur Rehman for previous work. The authors thank to LanPercept, a Marie Curie Initial Training Network funded through the 7th EU Framework Programme (316748).